# Creating multi-beam interference from two-beam interference with assistant of harmonics generation

**WUZHEN LI,**[1,2,3,*] **ZHIYUAN ZHOU,**[1,2,3,*,†] **LI CHEN,**[1,2,3] **YINHAI LI,**[1,2,3] **GUANGCAN GUO,**[1,2,3] **AND BAOSEN SHI**[1,2,3,‡]

[1]*CAS Key Laboratory of Quantum Information, University of Science and Technology of China, Hefei, Anhui 230026, China*
[2]*CAS Center for Excellence in Quantum Information and Quantum Physics, University of Science and Technology of China, Hefei 230026, China*
[3]*Hefei National Laboratory, University of Science and Technology of China, Hefei 230088, China*
*These authors contributed equally.*
[†]*zyzhou@ustc.edu.cn*
[‡]*drshi@ustc.edu.cn*

**Abstract:** Linear optics-based multi-beam interference (MBI), like the Fabry-Perot interferometer, plays an important role in precision optical metrology applications such as laser stabilization in optical clocks, precision spectroscopy, and gravitational wave detection. Here, we propose and experimentally verify a nonlinear optics-based MBI principle with the assistance of cascading and recycling harmonics generation of two-beam interference. By cascading and recycling the harmonics processes, in combining with optical power amplification (OPA) to compensate for power losses arising from limited nonlinear conversion efficiency, a total 16th harmonic is achieved, and the observed interference fringes gradually evolve from a sinusoidal curve to a Lorentz-like curve. In principle, there is no limitation on the number of cascading and recycling nonlinear processes with the assistance of OPAs and sharp interference fringes, analogous to those in a high-finesse cavity, can be obtained. The nonlinear optics-based MBI mechanism revealed here will find promising applications in precision optical metrology.

## 1. Introduction

Multi-beam interference (MBI) is a basic principle in optics. Typical optical elements or devices based on this principle are the Fabry-Perot interferometer or optical cavity [1-3]. The optical cavity is vitally important in modern optics; typical applications of the optical cavity are laser stabilization in optical clocks [4-7], precision spectroscopy [8-11], gravitational wave detection [12], etc. Nowadays, various techniques are developed to build an optical cavity for practical applications, and the optical cavity can be constructed with separated coating mirrors [13-16], monolithic optical materials with optical coatings [17, 18], and integrated micro-cavity which can be fabricated in large amount and small footprints with CMOS compatible platforms [19-21]. Two important parameters for an optical cavity are the free spectral ranges and finesse (or Q-factor). These two parameters need to be designed for a specific application scenario. The MBI-based devices mentioned above are all passive devices, and light is processed inside the devices with linear optical principles. Is it possible to realize other kinds of MBI based on nonlinear optics, which have properties similar to those of linear optics-based devices?

Our previous work [22] achieved a four-fold amplification of the relative phase between two polarization modes in a polarization interferometer through the cascaded second harmonic generation (SHG) process and proved that the phase amplification process is independent of the carrier frequency. This is because the relative phase between a superposed laser mode can be coherently inherited and combined in three-wave mixing processes, and the harmonics generated light contains multi-interaction paths between orthogonal superposed modes of the input fundamental beams. Based on this key principle, we reveal a new mechanism for generating MBI by cascading and recycling harmonics generation (HG) of a two-beam interference (TBI) signal from a Michelson interferometer (MI). To overcome the power losses in nonlinear processes due to less than unity conversion efficiency, an Erbium-doped fiber amplifier is adopted to compensate for the power losses of the harmonics generated beam. In the present configuration, a total 16th harmonics beam is generated. The observed interference fringes gradually evolve from a sinusoidal curve to a Lorentz-like curve by scanning the relative phase of the original MI. In principle, arbitrary orders of harmonics can be generated if sufficient copies of the basic recycling blocks are available. We can achieve a sharp peak after a certain number of HG, which is analog to a traditional high-finesse optical cavity.

## 2. Theory

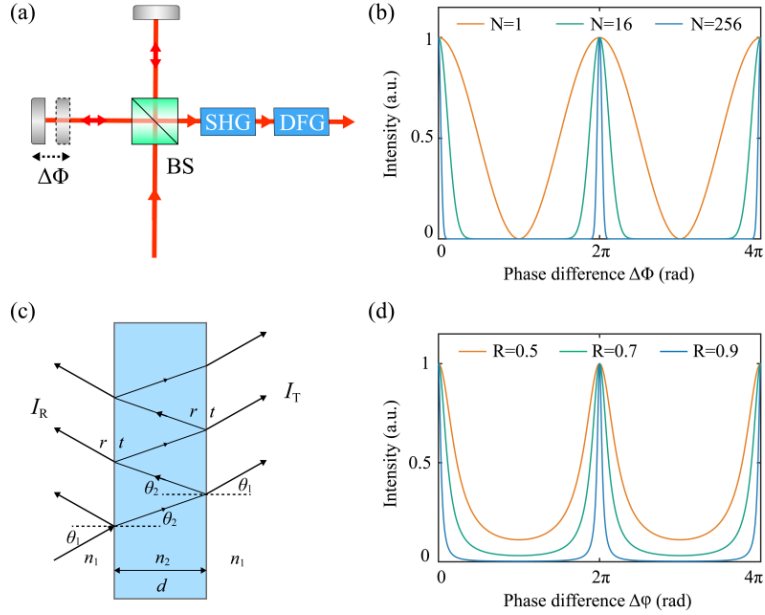

Fig. 1. Schematic diagram of MBI generation. (a) Michelson interferometer for generating MBI. (b) MBI fringes at different harmonics obtained by theoretical simulation. (c) Asymmetrical FP-cavity. (d) Theoretical transmission intensity of FP-cavity at different reflectances.

The general theoretical models will be given first. Figure 1 shows a graphical summary of the main concept of this work. As shown in Fig. 1(a), a SHG or N-th harmonic generation (NHG) module is placed after the MI. In wave optics, the interference field at the output ports of the MI can be expressed as:

$$E(\omega) = E_1(\omega) + e^{i\Delta\Phi} E_2(\omega), \quad (1)$$

where $E_1(\omega)$ and $E_2(\omega)$ are the amplitudes of the two superposed light fields at the beam splitter, and $\Delta\Phi$ is the phase difference between these two fields at the fundamental wavelength. After the SHG and NHG processes, the amplitudes of the generated second harmonic (SH) and the N-th harmonic can be expressed as [23-25]:

$$E(2\omega) \propto [E_1(\omega) + e^{i\Delta\Phi} E_2(\omega)]^2 = E_1(\omega)^2 [1 + \gamma e^{i\Delta\Phi}]^2,$$
$$E(N\omega) \propto [E_1(\omega) + e^{i\Delta\Phi} E_2(\omega)]^N = E_1(\omega)^N [1 + \gamma e^{i\Delta\Phi}]^N, \quad (2)$$

where $\gamma = E_2 / E_1$ is the ratio of the amplitude between the two original beams. The intensity of the SH and the N-th harmonic can be expressed as:

$$I(2\omega) = |E(2\omega)|^2 \propto I_1(\omega)^2[1+\gamma^2+2\gamma\cos(\Delta\Phi)]^2,$$
$$I(N\omega) = |E(N\omega)|^2 \propto I_1(\omega)^N[1+\gamma^2+2\gamma\cos(\Delta\Phi)]^N. \quad (3)$$

For a symmetrical MI, in which the amplitude ratio between the two beams is 1, equation (3) can be further simplified as:

$$I(2\omega) \propto 16I(\omega)^2 \cos^4(\frac{\Delta\Phi}{2}),$$
$$I(N\omega) \propto 4^N I(\omega)^N \cos^{2N}(\frac{\Delta\Phi}{2}). \quad (4)$$

The intensity fringes of the different harmonics are simulated in Fig. 1(b), we can see that through the HG process, we can directly generate multi-beam like interference fringes from TBI. For the linear optics-based MBI, such as a symmetrical FP-cavity shown in Fig. 1(c), the amplitude and intensity of the transmitted beam from the FP-cavity can be expressed as:

$$E_T = \frac{(1-r^2)e^{i\Delta\varphi/2}}{1-r^2 e^{i\Delta\varphi}} = (1-r^2)e^{i\Delta\varphi/2}\sum_{n=0}^{\infty}(r^2 e^{i\Delta\varphi})^n,$$
$$I_T = |E_T|^2 = \frac{(1-r^2)^2}{1+r^4-2r^2\cos(\Delta\varphi)} = \frac{T^2}{1+R^2-2R\cos(\Delta\varphi)}, \quad (5)$$

where $\Delta\varphi$ denotes the phase difference between adjacent transmitted beams, $r$ is the amplitude reflection coefficient, $R=|r|^2$ is the intensity reflectance, $T=1-R$ is the intensity transmittance. The simulated curves of the traditional FP-cavity at different reflectances are shown in Fig. 1(d). By comparing Eqs. (2), (3), (4) with (5), as well as the simulated fringes, we can see that a traditional FP-cavity is a superposition of infinite multi-beams with decreasing coefficient, and the MBI based on the HG process is a superposition finite multi-beams with a binomial coefficient distribution.

To clearly quantify the MBI effect caused by the HG process, we introduce the finesse in analog to a traditional FP-cavity, a crucial parameter for evaluating the phase resolution capability of the interferometer, defined as:

$$F = \frac{2\pi}{\delta}, \quad (6)$$

where $\delta$ denotes the full width at half maximum of the fringes. Therefore, the finesse of the interference fringes after the NHG process can be expressed as:

$$F_N = \frac{\pi}{2\arccos(2^{-\frac{1}{2N}})}. \quad (7)$$

Equation (7) demonstrates that the finesse of the TBI fringes is significantly improved after the NHG process, which also indicates that the phase sensitivity and resolution of the interferometer are enhanced.

Similar to the phase amplification process proposed in our previous work [22], the process of generating MBI from TBI is also independent of the carrier frequency. The schematic diagram is shown in Fig. 1(a); a DFG module is inserted after the SHG module, which can convert the frequency of the second harmonic from $2\omega$ back to $\omega$. If the pump power is undepleted in the DFG process, the resulting fringe intensity can be expressed as:

$$I_{DFG}(\omega) \propto I_{SHG}(2\omega)I_{pump}(\omega)$$
$$= I_1(\omega)^2[1+\gamma^2+2\gamma\cos(\Delta\Phi)]^2 I_{pump}(\omega). \quad (8)$$

where $I_{pump}(\omega)$ is a constant, denoting the pump intensity. Equation (8) demonstrates that MBI can still be generated, even without a change in frequency. Moreover, if the nonlinear conversion efficiencies for both SHG and DFG are high enough, the cascaded SHG-DFG processes can be cycled more times to obtain higher finesse. Fortunately, the scheme becomes feasible by selecting appropriate nonlinear crystals, optimizing the beam waist radius after focusing, and using existing mature optical power amplification technology to amplify the fringe power after each cycle. More details about the recycling scheme are described in the Supplement 1.

## 3. Results

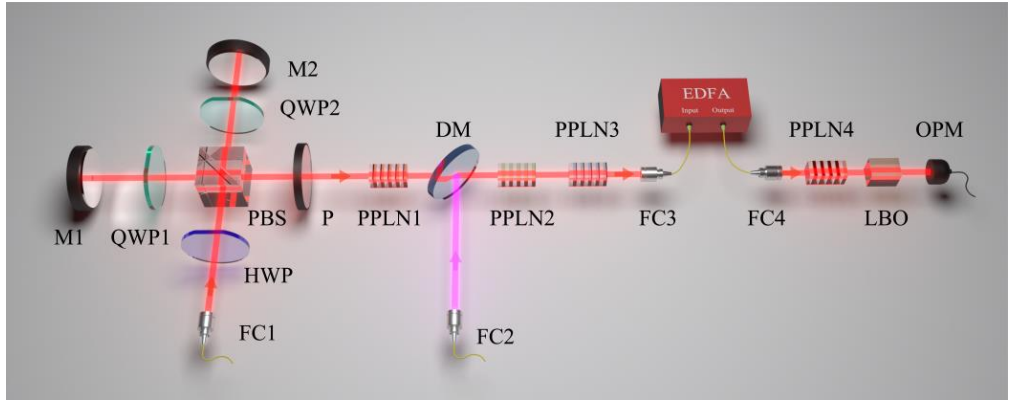

Fig. 2. Experimental setup schematic. PBS, polarizing beam splitter; HWP, half-wave plate; QWP, quarter-wave plate; M, mirror; DM, dichroic mirror; PZT, piezoelectric transducer; T, translation; P, polarizer; PPLN, periodically poled lithium niobate crystal; EDFA, Er-doped fiber amplifier; FC, fiber collimators; LBO, lithium triborate crystal; OPM, optical power meter.

We now experimentally demonstrate the predictions described in the theory above. As shown in Fig. 2, the 1560 nm pulsed light with a pulse duration of 212 ps emitted from the fiber collimator (FC1) serves as the input signal for the MI. The resulting TBI fringes are shown in Fig. 3(a), exhibiting a finesse of 2. The y-axis in Fig. 3 represents the measured optical power, and the x-axis represents the change in the optical path difference (OPD). The 1560 nm TBI fringes first pass through a type-0 periodically poled lithium niobate (PPLN) crystal with a period of 19.62 μm and a length of 25 mm to generate a 780 nm SH as shown in Fig. 3(b). Then, the 780nm SH is combined with the 1040 nm synchronous pump pulse to generate 3120 nm mid-infrared (MIR) interference fringes through DFG in a type-0 PPLN crystal with a period of 21.35 μm and a length of 30 mm, and the result is shown in Fig. 3(c). The MIR interference fringes are then converted back to the 1560 nm through the second SHG process. The crystal used to generate 1560 nm SH shown in Fig. 3(d) is a type-0 PPLN crystal with a period of 34.8 μm and a length of 25 mm. The measured interference fringe finesses of the 780 nm SH, MIR, and 1560 SH are 2.6, 2.6, and 3.55, respectively, which are slightly lower than the theoretical values of 2.747, 2.747, and 3.83. As described in the Supplement 1, this is primarily attributed to two factors: on the one hand, the maximum conversion efficiency of the 780 nm SHG process is about 65.2%, resulting in the 1560 nm signal no longer strictly obeying the undepleted pump approximation condition; on the other hand, the spectral bandwidth of the 1560 nm signal is broadened after the SHG process, and the filtering effect induced by the nonlinear process leads to a reduction in finesse. A comparison of the finesses of the 780 nm SH and MIR interference fringes demonstrates that the DFG process does not change the finesse, which is consistent with theoretical predictions. After passing through the cycle module consisting of the cascaded SHG, DFG, and SHG processes, the wavelength of the interference fringes returns to 1560 nm after multiple changes, but the finesse is improved from 2 to 3.55, which demonstrates that the MBI based on nonlinear processes is independent of frequency.

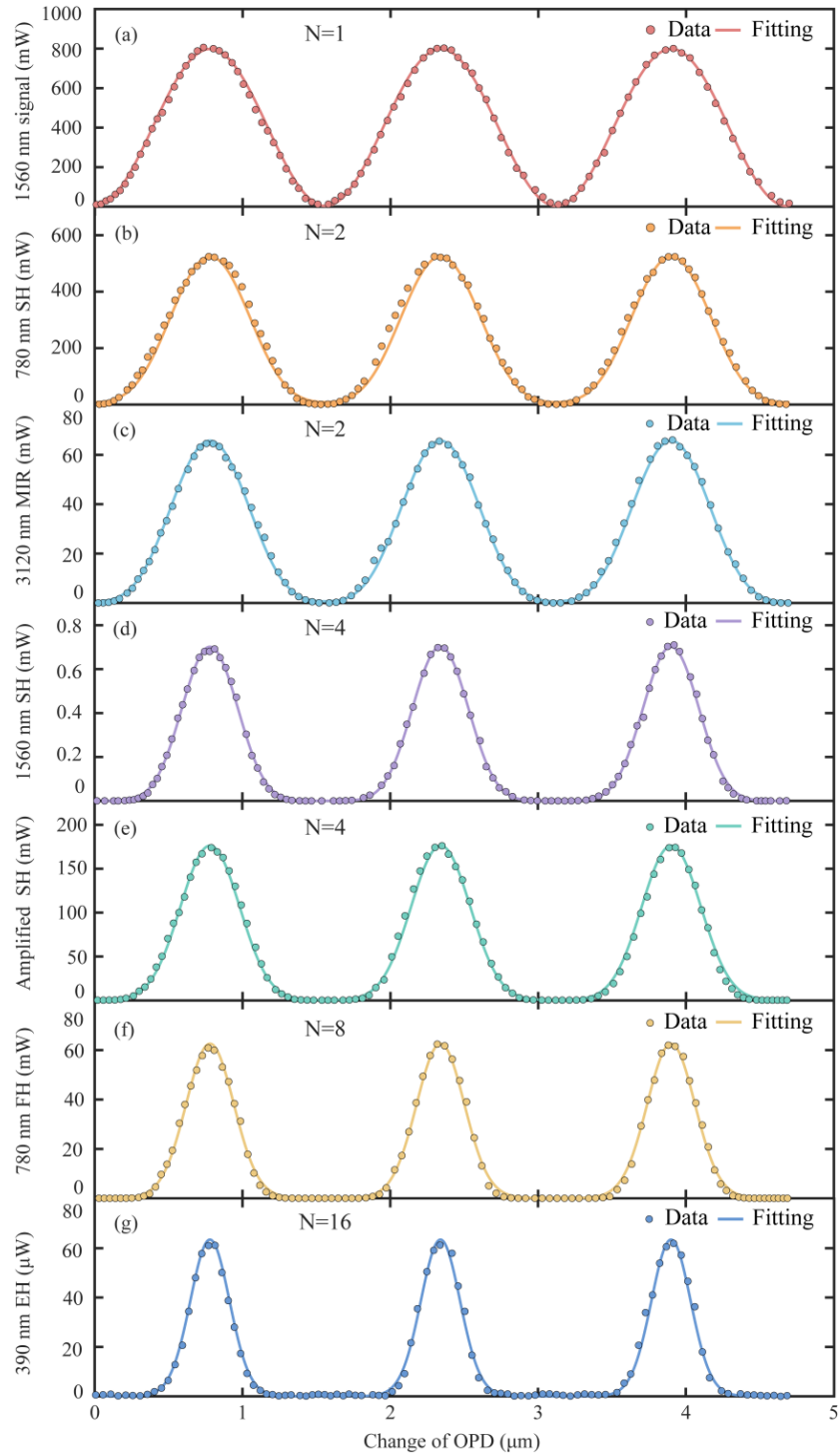

Fig. 3. Measured interference fringes of (a) 1560 nm signal light, (b) 780 nm SH, (c) 3120 nm MIR, (d) 1560 nm SH, (e) amplified 1560 nm SH, (f) 780 nm FH, and (g) 390 nm EH. N represents the number of harmonic orders required to directly generate the equivalent interference fringes, as shown in Eq. (4).

Although the interference fringes become sharper after one cycle, the output power of the 1560 nm SH is insufficient for the next cycle due to the low conversion efficiency of the SHG-DFG-SHG cycle module. Fortunately, mature optical amplification techniques, such as doped fiber amplifiers and semiconductor optical amplifiers, can be employed to amplify the power of interference fringes, thereby increasing the number of cycles. In the experiment, a homemade Er-doped fiber amplifier (EDFA), shown in Fig. 2, is used to amplify the 1560 nm SH. As shown in Fig. 3(e), the finesse of the interference fringes after amplification is slightly reduced to 3.4, which is attributed to the signal distortion caused by the nonlinear gain of the amplifier. The average gain of the input signal in the experiment is approximately 24 dB, which is sufficient for the next cycle. Furthermore, the gain and linearity of the amplifier can be further improved by optimizing the gain medium, pump power, and structure of the amplifier. The amplified 1560 nm SH interference fringes then pass through a cascaded SHG module consisting of a PPLN crystal and a lithium triborate (LBO) crystal to generate a 780 nm fourth harmonic (FH) and a 390 nm eighth harmonic (EH), as shown in Figs. 3(f) and 3(g). The fringe finesses of the 780 nm FH and 390 nm EH are 4.1 and 5.2, respectively. In the cascaded SHG module, the nonlinear efficiency of the first SHG process is ~ 36.5%, while that of the second SHG process is only ~ 0.1%. This is mainly because the intensity of the interference fringes is significantly reduced after the first SHG process, and the nonlinear coefficient of the LBO crystal is smaller than that of the PPLN crystal. In addition, for the SHG process based on the LBO crystal, the walk-off effect also leads to a decrease in conversion efficiency. More information about the experimental setup is given in Supplement 1.

**4. Conclusions**

In summary, a full theoretical and experimental description of the nonlinear optics-based MBI is given in this Letter. In the experiment, we obtain MBI fringes with a finesse of 5.2 from the TBI by cascading and recycling HG processes. The enhancement in the sharpness of the interference fringes indicates that even subtle phase shifts can cause noticeable changes in the fringe pattern. Therefore, through cascading and recycling nonlinear processes, the two-beam interferometer acquires many of the advantages of the multi-beam interferometer, such as enhanced phase sensitivity and resolution. In addition, the frequency independence of the nonlinear optics-based MBI is verified by a specially designed SHG-DFG-SHG sandwich structure. In our experiment, the power conversion efficiency of the SHG-DFG-SHG cycle module is less than 0.1%, which is mainly attributed to two factors: first, due to energy conservation and quantum loss, the power conversion efficiency of the DFG process is about 12.5%; second, the MIR power entering the second SHG process is low, resulting in a conversion efficiency of only 1.1%. Although the power of the interference fringes is significantly reduced after each cycle, it can be amplified by a linear optical power amplifier, as demonstrated in our experiment. Therefore, the SHG-DFG-SHG cycle module can be replicated infinitely in principle using linear optical amplifiers for power compensation. Currently, the state-of-the-art high-finesse cavity has a finesse of around 1.9 million [26]; if we can achieve 20 cycles of SHG-DFG-SHG processes, the finesse of the nonlinear optics-based MBI can surpass the traditional techniques. This may beat the present limit of coating-based linear optics devices. The present scheme may also be combined with high harmonics generation in gas and solids [24, 25], which may simplify the whole experimental setup.

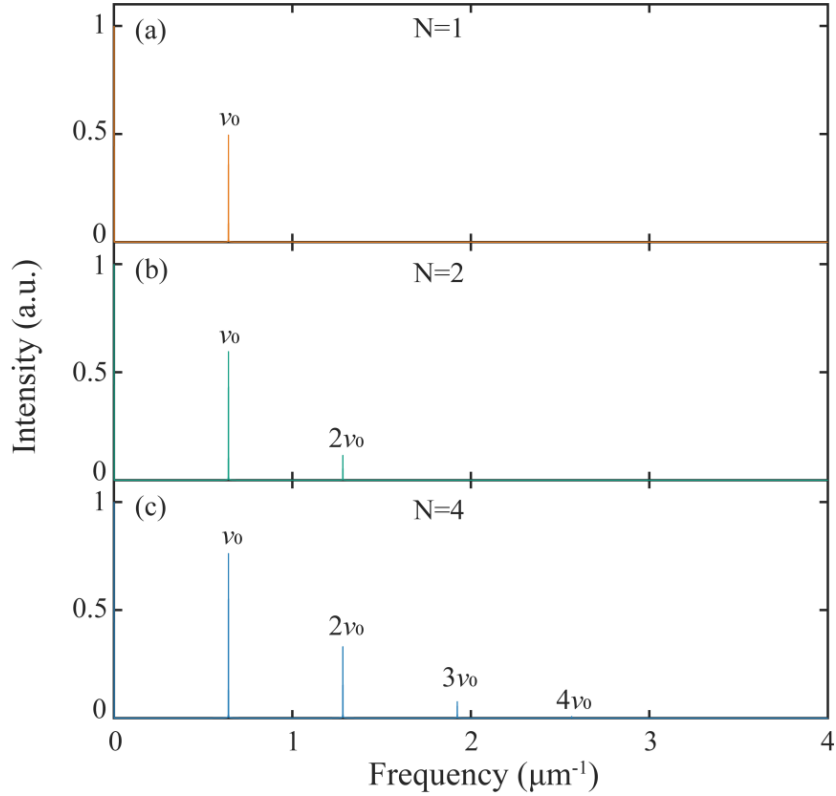

Fig. 4. Power spectra of (a) 1560 nm signal, (b) 780 nm SH, and (c) 1560 nm SH after Fourier transformation. N represents the number of harmonic orders, and v0 denotes the fundamental frequency of the TBI fringes.

We believe that the MBI based on nonlinear processes proposed in this Letter will be applied to many precision measurement scenarios based on TBI, such as the measurements of the optical properties (for example, dispersion and absorption coefficients) of transparent materials, other quantities such as displacements, angles, and electrical and magnetic fields, which can be transduced into changes in an optical path length, or interference-based imaging for enhanced image resolution, etc. Furthermore, as shown in Eq. (2), the interference fringes generated after the HG process contain additional frequency information, which can be extracted through filtering to obtain the desired high-frequency components. To clarify this, we performed a Fourier transform on the fitting curves of 1560 nm signal, 780 nm SH, and 1560 nm SH shown in Fig. 3, and the resulting power spectra are shown in Fig. 4. As the harmonic order increases, the power spectrum of the interference fringes exhibits more high-frequency components which also indicates that the phase resolution capability of the interferometer is enhanced. In addition, in the cascading and recycling nonlinear processes, the wavelength of the interference fringes undergoes multiple changes, so we can flexibly design the working wavelength and detection wavelength of the interferometer according to specific needs. The work presented here will revolutionize our understanding of MBI and may open new avenues for interference-based precision metrology.

**Funding.** National Key Research and Development Program of China (2022YFB3903102, 2022YFB3607700), National Natural Science Foundation of China (NSFC) (62435018), Innovation Program for Quantum Science and Technology (2021ZD0301100), USTC Research Funds of the Double First-Class Initiative, and Research Cooperation Fund of SAST, CASC (SAST2022-075).

**Acknowledgment.** We sincerely thank Xiaohua Wang, He Zhang and Mengyu Xie for their help in this work.

**Disclosures.** The authors declare no conflicts of interest.

**Data availability.** Data underlying the results presented in this paper are not publicly available at this time but may be obtained from the authors upon reasonable request.

**Supplemental document.** See Supplement 1 for supporting content.

# CREATING MULTI-BEAM INTERFERENCE FROM TWO-BEAM INTERFERENCE WITH ASSISTANT OF HARMONICS GENERATION: SUPPLEMENTAL DOCUMENT

WUZHEN LI,[1,2,3,*] ZHIYUAN ZHOU,[1,2,3,*,†] LI CHEN,[1,2,3] YINHAI LI,[1,2,3] GUANGCAN GUO,[1,2,3] AND BAOSEN SHI[1,2,3,‡]

[1]*CAS Key Laboratory of Quantum Information, University of Science and Technology of China, Hefei, Anhui 230026, China*
[2]*CAS Center for Excellence in Quantum Information and Quantum Physics, University of Science and Technology of China, Hefei 230026, China*
[3]*Hefei National Laboratory, University of Science and Technology of China, Hefei 230088, China*
*\*These authors contributed equally.*
*†zyzhou@ustc.edu.cn*
*‡drshi@ustc.edu.cn*

## S1. Experimental setup

The detailed schematic of the experimental setup is shown in Fig. S1. The Er-doped fiber laser (EDFL) is a homemade polarization-maintaining mode-locked fiber laser with a repetition rate of 21.66 MHz. The pulse signal is then amplified by a homemade Er-doped fiber amplifier (EDFA1), and the maximum output power is close to 2 W. The optical signal from a 10% tap-coupler (TAP) is detected by a 10 GHz photodiode detector (PD), where it is converted into an electrical pulse signal. This electrical signal is subsequently amplified by an RF amplifier and used to drive a 10 GHz electro-optical intensity modulator (EOM). As a result, the continuous-wave light emitted by the 1040 nm continuously tunable diode laser (CTL) is modulated into synchronous pulsed light with the same repetition frequency. The DC-B voltage can be adjusted to optimize the pulse extinction ratio. The Yb-doped fiber amplifier (YDFA) is a homemade, multi-stage, polarization-maintaining fiber amplifier that can amplify the power of the 1040 nm modulated pulse to over 5 W.

We define the 1560 nm pulse light from EDFA1 as the signal and the 1040 nm synchronous pulse light from YDFA as the pump. The linearly polarized signal is first converted into a 45°-polarized beam using a half-wave plate (HWP) and then split into two orthogonally polarized beams of equal intensity by the polarizing beam splitter (PBS). In each arm of the Michelson interferometer, a setup consisting of a mirror and a quarter-wave plate (QWP) is used to convert the polarization state of the signal into its orthogonal state. Mirror M1 is fixed on a translation stage that is used to perform rough adjustments of the optical path difference (OPD) between the two arms. Mirror M2 is fixed on a piezoelectric transducer (PZT) that is used to enable fine adjustment of the OPD within a 4.7 μm range; the PZT is driven using an amplified triangular wave signal. The 1560 nm interference fringes, generated after passing through the polarizer (P), are converted to 780 nm via second harmonic generation (SHG) in a type-0 periodically poled lithium niobate (PPLN1) crystal with a length of 25 mm and a poling period of 19.62 μm, and the operation temperature of the PPLN1 crystal is set to be 44.5 °C to fulfill the quasi-phase-matching condition. Then, the 780 nm second harmonic (SH) and 1040 nm pump pulse are spatially combined by a dichroic mirror (DM) before being focused into another type-0 PPLN2 crystal with a length of 30 mm, a poling period of 21.35 μm and the quasi-phase-match temperature of 38.2 °C, which facilitates the difference frequency generation (DFG) to obtain the MIR interference fringes at 3120 nm. An optical delay line (ODL) is inserted in the path to optimize the temporal overlap of the two interacting pulses.

A type-0 PPLN3 crystal with a length of 25 mm is used to convert the MIR interference fringes back to 1560 nm by the SHG process. The poling period is chosen to be 34.8 μm, and the operation temperature is stabilized at 27 °C. The 1560nm SH then enters the homemade Er-doped fiber amplifier (EDFA2) for power amplification, and a 200 GHz bandwidth dense wavelength division multiplexer (DWDM), channel 22 with a central wavelength of 1559.79 nm, is used to filter out the ASE noise generated by the amplifier. The amplified 1560 nm SH first passes through the PPLN4 crystal with the same specifications as the PPLN1 crystal to generate the 780 nm fourth harmonic (FH) and finally passes through a lithium triborate (LBO) crystal with a length of 20 mm and a cutting angle of 33.7° to generate the 390 nm eighth harmonic (EH). M3-M8 are used to separate the interference fringes from the optical path, after which the interference fringes are detected and recorded using optical power meters. In Table S1, we list the relevant specifications of the involved lenses and filters.

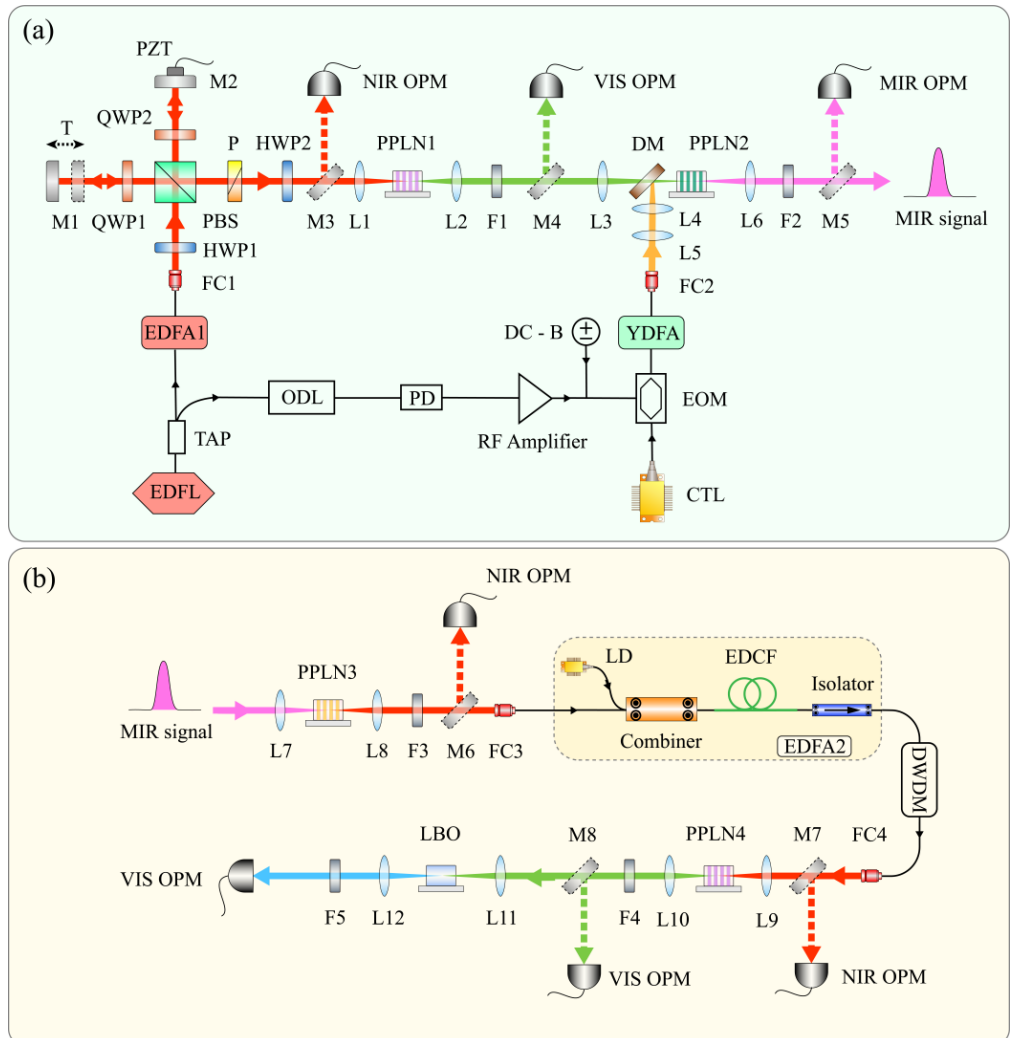

Fig. S1. Detailed schematic for the experimental setup. EDFL: Erbium-doped mode-locked fiber laser; EDFA: Er-doped fiber amplifier; ODL: optical delay line; PD: photodiode detector; CTL: continuously tunable diode laser; EOM: electro-optical intensity modulator; YDFA: Yb-doped fiber amplifier; FC: fiber collimator; HWP: half-wave plate; QWP: quarte-wave plate; M: mirror; T: Translation Stage; PZT: piezoelectric transducer; PBS: polarization beam splitter; P: polarizer; L: lens; F: filter; DM: dichroic mirror; PPLN: periodically poled lithium niobate crystal; LBO: lithium triborate crystal; LD: laser diode; EDCF: Er-doped double-clad fiber; DWDM: dense wavelength division multiplexer; NIR (VIS, MIR) OPM: near-infrared (visible, mid-infrared) optical power meter.

TABLE S1. Specification for the involved lens and spectral filters

| Symbol | Item | Parameters |
|---|---|---|
| L1 | K9 Plano-Convex Spherical Lens | F = 100 nm; AR coating: 1200-1600 nm |
| L2 | K9 Plano-Convex Spherical Lens | F = 100 nm; AR coating: 700-1100 nm |
| L3 | K9 Plano-Convex Spherical Lens | F = 100 nm; AR coating: 700-1100 nm |
| L4 | K9 Plano-Convex Spherical Lens | F = 75 nm; AR coating: 700-1100 nm |
| L5 | K9 Plano-Convex Spherical Lens | F = 75 nm; AR coating: 700-1100 nm |
| L6 | $CaF_2$ Lens | F = 75 nm |
| L7 | $CaF_2$ Lens | F = 75 nm |
| L8 | K9 Plano-Convex Spherical Lens | F = 100 nm; AR coating: 1200-1600 nm |
| L9 | K9 Plano-Convex Spherical Lens | F = 75 nm; AR coating: 1200-1600 nm |
| L10 | K9 Plano-Convex Spherical Lens | F = 100 nm; AR coating: 700-1100 nm |
| L11 | K9 Plano-Convex Spherical Lens | F = 75 nm; AR coating: 700-1100 nm |
| L12 | N-BK7 Plano-Convex Spherical Lens | F = 50 nm; AR coating: 350-700 nm |
| F1 | Low-pass filter | Cut-off wavelength: 1000 nm |
| F2 | Dichroic mirror 1 Dichroic mirror 2 | Dichroic mirror 1: R 1040 nm, AR 1560 & 3120 nm; Dichroic mirror 2: R 780 nm, AR 3100 nm |
| F3 | Band-pass filter | CWL: 1560 nm FWHM: 20 nm |
| F4 | Low-pass filter | Cut-off wavelength: 1000 nm |
| F5 | Band-pass filter | CWL: 390 nm FWHM: 13 nm |

CWL: center wavelength; FWHM: full width at half maximum.

S2. Spectro-temporal characteristics of optical pulses

In our experiment, the nonlinear frequency conversion is implemented based on the coincident pulse pumping. To further optimize the conversion performance, we have carefully engineered the spectro-temporal properties of the involved optical pulses and precisely controlled the temporal overlap of the 1040 nm pump pulse and the 780 nm SH by adjusting the ODL. Figure S2(a) illustrates synchronized pulse trains from the YDFA and EDFA, which are utilized in subsequent SHG and DFG processes. As shown in Figs. S2(b) and (c), the spectral bandwidth of the 1040 nm pump pulse from the YDFA is about 0.032 nm, and the pulse duration measured by a 25 GHz high-speed photodetector and a 12.5 GHz oscilloscope is 1.35 ns. The optical spectrum of the 1560 nm signal pulse from the EDFA is given in Fig. S2(d), which shows a narrow bandwidth of 0.072 nm, and the corresponding pulse duration is 212 ps, as shown in Fig. S2(e). The shorter pulse delay of the signal relative to the pump not only facilitates higher frequency doubling efficiency but also improves the conversion efficiency of the 780 nm SH in the DFG process. In the spectral domain, both the signal and pump sources have narrow bandwidths, which allows them to approach the phase-matching bandwidth.

Using the Word styles. In this template, styles such as "01 Title" for the manuscript appear in the MS Word Styles ribbon (or toolbar for older versions of Word). Apply the appropriate style before typing or apply the style to existing text. It is possible to paste manuscript text into this template or attach this template to an existing manuscript.

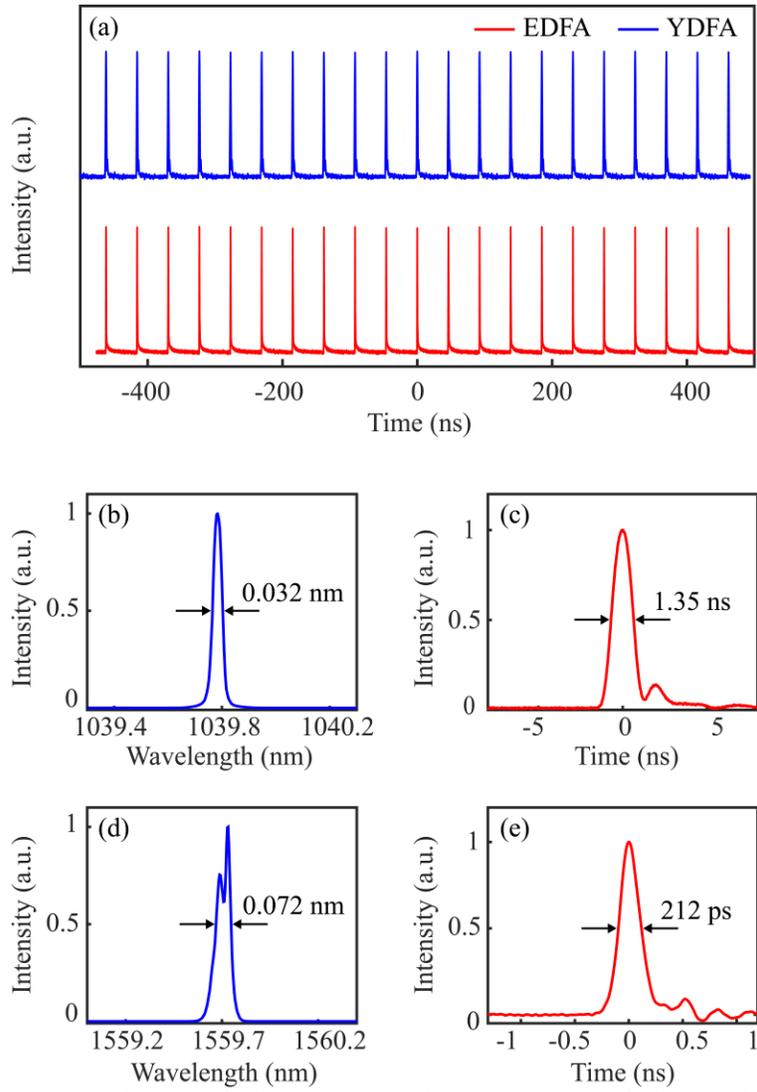

Fig. S2. Spectro-temporal characterization for the involved optical pulses. (a) Pulse trains from synchronized EDFA and YDFA. (b)-(c) The spectral and temporal intensity distribution of the 1040 nm pump pulse. (d)-(e) The spectral and temporal intensity distribution of the 1560 nm signal pulse.

## S3. Data acquisition and processing

Figure S3 shows the raw data recorded by an optical power meter (OPM) at different harmonics. We first use the mirror M3 in Fig. S1 to separate the 1560 nm signal interference fringes and then record them using a near-infrared (NIR) OPM. The measured optical power of the interference fringes against the PZT scanning voltage is shown in Fig. S3(a). In the scanning voltage range of 0 to 120 V, the optical power varies by ~3 fringe periods, indicating a change of ~ 4.68 μm in the optical path difference (OPD) between the two arms of the interferometer. After recording the interference fringes of the 1560 nm signal, we removed M3 to perform the SHG experiment. We use the M4 and a visible (VIS) OPM to measure the interference fringes of 780nm SH, and the corresponding result is given in Fig. S3(b). Then, we measured the interference fringes of MIR, 1560 nm SH, amplified 1560 nm SH, 780 nm FH, and 390 nm EH; the results are shown in Figs. S3(c)-(g).

Due to the nonlinear dependence of the displacement on the scanning voltage in the PZT, the fringe periods are non-uniformly distributed with respect to the scanning voltage. In order to unify the period of interference fringes, we need to convert the scanning voltages into changes in OPD, as shown in Fig. 3 in the main text. We first perform function fitting on the experimental data in Fig. S3 to obtain an accurate function relationship between changes in OPD and voltages, and then we make use of the function to convert the scanning voltages into the corresponding changes in OPD.

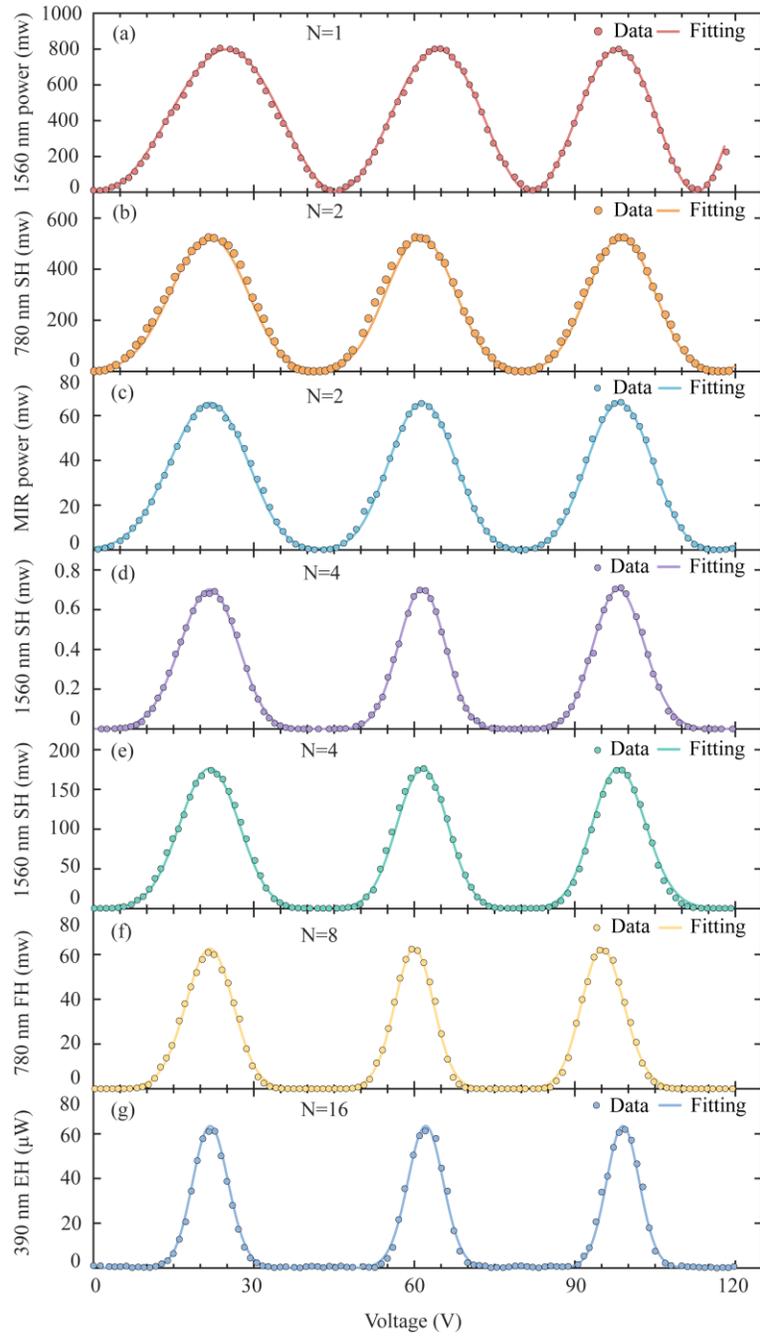

Fig S3. Interference fringes versus PZT scanning voltage for the 1560 nm signal (a), 780 nm SH (b), 3120 nm MIR (c), 1560 nm SH (d), amplified 1560 nm SH (e), 780 nm FH (f) and 390 nm EH (g). The dots represent the measured raw data points, while the solid lines correspond to the fitted curves.

## S4. Scheme for achieving higher finesse

Although the highest interference fringe finesse we obtained in the experiment is 5.2, higher fringe finesse can be achieved by a scheme, as shown in Fig. S4(a). Any variation in physical quantities that affect the optical path length is sensed by the two-beam interferometer, causing intensity changes in the interference pattern. After the harmonic generation process, the sinusoidal interference curve evolves into a Lorentz-like curve. By increasing the harmonic order, the finesse of the interference fringes will also increase. To improve the order of harmonic waves, one can cascade more SHG modules in addition to using high harmonic generation materials. For second-order nonlinear crystals, the complex high-order nonlinear processes are relatively weak; therefore, the noise is easy to filter. As shown in Fig. S4(a), we can cascade multiple SHG-DFG-SHG-EDFA cycle modules after the two-beam interferometer. The advantage of this cycle module is that the high-efficiency DFG of type 0 quasi-phase-matching can be utilized. For the SHG-DFG recycling scheme, the DFG should be based on the type-II phase-matching condition, where the efficiency is much lower than that of type-0 phase-matching. The EDFA is used to compensate for the power loss caused by the nonlinear conversion process. In principle, by employing optical power amplifiers for power compensation, there is no limitation on the number of cycles for the SHG-DFG-SHG module, and sharp interference fringes, analogous to those in high-finesse cavities, can be obtained. As shown in Fig. S4(b), the finesse of the nonlinear optics-based MBI increases exponentially with the number of cycles of the SHG-DFG-SHG module, and with only 20 cycles, the finesse of the nonlinear optics-based MBI can exceed the current maximum finesse of 1.9 million achieved by traditional FP-cavity. Moreover, the input or output light can not only work at the fundamental frequency $\omega$ but also at $2\omega$ or $\omega/2$, which can be flexibly designed according to the working band of the interferometer and the detector. We should point out that the construction of the cycle module is diverse and can be flexibly designed based on specific requirements and experimental conditions. In addition to the SHG-DFG-SHG structure, the cycle module can also be designed as SHG-SHG-DFG, among others.

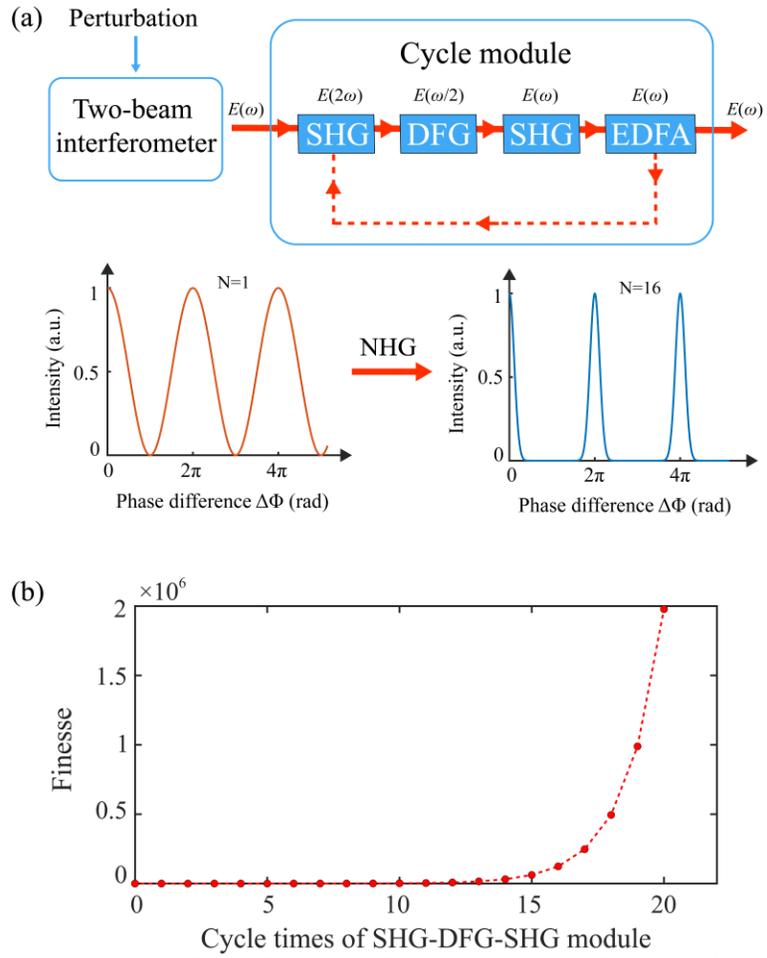

Fig. S4. (a) Schematic diagram of the SHG-DFG-SHG-EDFA cycle module. (b) The finesse of nonlinear optics-based MBI varies with the number of cycles of the SHG-DFG-SHG module.

S5. Distortion of interference fringes

In our experiment, the obtained finesses of the interference fringes are lower than the theoretical value, which can be attributed to two parts: one is the signal distortion caused by the nonlinear frequency conversion process, and the other is the signal distortion caused by the amplifier. For the SHG process, when the conversion efficiency is high, the incident pump power will not satisfy the undepleted approximation, and Eq. (2) in the main text will no longer satisfy. In this case, the amplitude and intensity of the SH can be expressed as [23]

$$E_{SHG}(2\omega) \propto E(\omega)\tanh[\alpha E(\omega)],$$
$$I_{SHG}(2\omega) \propto I(\omega)\tanh^2[\alpha\sqrt{I(\omega)}],$$
(S1)

where $\alpha$ denotes the effective length coefficient. If the interference fringes entirely violate the pump undepleted approximation, our theoretical analysis yields a lower limit for the finesse of 2.3. In the experiment, we characterized the power efficiency of 1560 nm signal light to generate 780 nm SH through the SHG process. The results are shown in Figs. 5S(a) and (b), where the red dashed line is the fitting curve obtained according to Eq. (S1). The conversion efficiency of the SHG process initially increases linearly with the incident signal light power, during which Eq. (2) in the main text holds true; however, as the signal power continues to increase, the conversion efficiency gradually saturates, and Eq. (2) is no longer applicable. In addition, we also measured the power efficiency and quantum conversion efficiency of the DFG process. At the fixed pump power of 3.5 W, the linear dependence of the output MIR power on the incident 780 nm SH power is validated as given in Fig. S5(c). The obtained power conversion efficiency is 16.1%, and the corresponding quantum efficiency is 64.4%. Figure S5(d) shows the dependence of the generated MIR power on the input pump power at a fixed 780 nm SH power of 1 W. When the pump power reaches 2.5W, the MIR power tends to saturate, and the corresponding maximum QCE is 63%. Therefore, by increasing the pump power to satisfy the undepleted approximation condition, the conversion efficiency of the DFG can be optimized without affecting the finesse of the interference fringes.

As we all know, the spectral bandwidth of the narrow-bandwidth light source will be broadened during the SHG and SFG processes. As shown in Fig. S6, after one SHG-DFG-SHG cycle module, the spectral bandwidth of the 1560 nm signal is broadened from the initial 0.072 nm to 0.316 nm. However, due to the phase mismatch of the nonlinear process, the acceptance bandwidth of the nonlinear crystal is limited, which causes spectral filtering of the incident light, thereby affecting the finesse of the interference fringes.

Another factor that causes the reduction of fringe finesse in the experiment is the amplifier. As shown in Fig. S1(b), the amplifier EDFA2 used in the experiment mainly consists of a beam combiner, a 3.5 m long erbium-doped double-clad fiber, and a high-power isolator. In addition, a narrowband filter is essential. Otherwise, the output interference fringes will contain a high background noise, as shown in Fig. S7(a). Therefore, we inserted a DWDM with a bandwidth of 200 GHz after the amplifier to filter out the background noise caused by ASE. Then, we characterized the small signal gain characteristics of the amplifier, and the result is shown in Fig. S7(b). When the input signal power is lower than 0.8 mW, the output power of the amplifier depends linearly on the input power, and the gain of the amplifier obtained from the slope of the linear fitting is 24 dB. However, as the input signal optical power further increases, the gain gradually decreases and eventually reaches gain saturation. Therefore, to ensure the amplified signal remains undistorted, the input signal power must be maintained within the linear operating region, far from gain saturation. We set the peak power of the input 1560 nm SH interference fringes below 0.8 mW and measured the spectrum of the amplifier output as shown in Fig. S7(c). The spectrum of the amplified interference fringes is slightly distorted due to the inhomogeneity of the fiber gain spectrum and the nonlinear effect during the amplification process. Comparing the interference fringes before and after amplification, as shown in Fig. S7(d), it can be seen that the interference fringes after amplification are slightly wider than those before amplification, which is mainly because the amplifier does not strictly work in the linear region. The gain and linearity of the amplifier can be improved by further optimizing the pump power, incident signal light power, gain fiber length, and amplifier structure (such as multi-stage amplification).

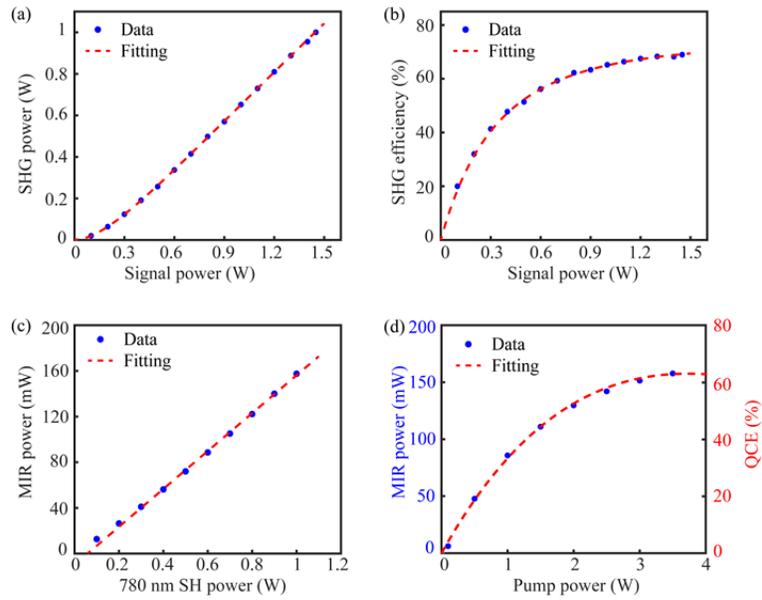

Fig. S5. (a) 780 nm SH power as a function of 1560 nm signal power. (b) The conversion efficiency of the SHG process against the incident signal power. (c) Linear dependence between the input 780 nm SH power and the MIR power. (d) MIR power and QCE are functions of input pump power.

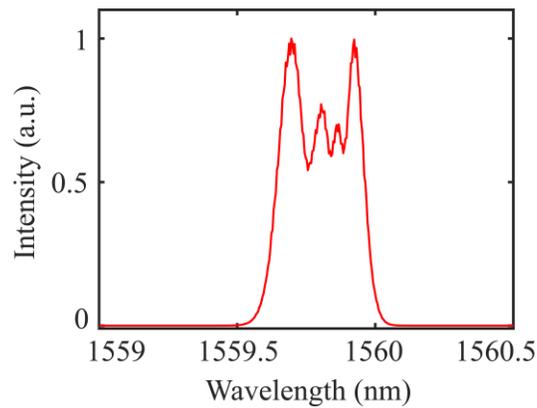

Fig. S6. Spectrum of interference fringes generated after passing through the SHG-DFG-SHG cycle module.

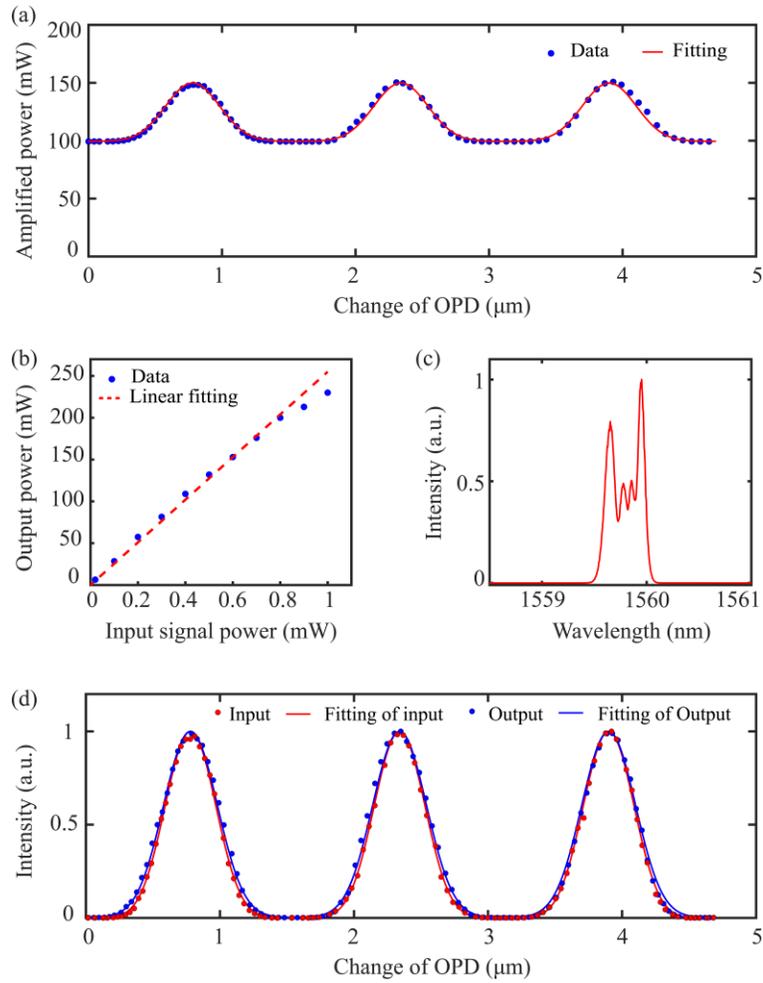

Fig. S7. (a) Interference fringes at the amplifier output without DWDM. (b) Amplifier output power versus input signal power for 6.5 W pump power. (c) The spectrum of interference fringes after power amplification. (d) Input (red) and output (blue) interference fringes of the amplifier. The dots represent experimental data points, and the solid lines are the theoretical fitting curves.